\documentclass[sigconf, table, dvipsnames]{acmart}

\usepackage{listings}
\usepackage{booktabs}
\usepackage{array}
\usepackage{multirow}
\usepackage{enumitem}
\usepackage{lineno}
\usepackage{subcaption}
\usepackage{tikz}
\usepackage{tabularx}
\usepackage{placeins}
\usepackage{pdflscape}
\usepackage{float}
\usepackage{amsmath}

\usepackage{algorithm}
\usepackage{algpseudocode}

\usetikzlibrary{
  positioning,
  arrows.meta,
  shapes.geometric,
  fit,
  backgrounds,
  calc
}

\DeclareUnicodeCharacter{2264}{\ensuremath{\leq}}

\renewcommand\footnotetextcopyrightpermission[1]{}

\AtBeginDocument{%
  \providecommand\BibTeX{{%
    \normalfont B\kern-0.5em{\scshape i\kern-0.25em b}\kern-0.8em\TeX}}}

\graphicspath{{./}}

\begin{document}

\title{Fairness-Aware Multimodal Transformer Modeling for Real-Time Student Attention Estimation}

\author{Christoforos Fragkiadakis}
\affiliation{%
  \institution{University of Amsterdam}
  \city{Amsterdam}
  \country{The Netherlands}
}
\email{christoforos.fragkiadakis@student.uva.nl}

\author{Seyed Sahand Mohammadi Ziabari}
\authornote{Corresponding author.}
\affiliation{%
  \institution{University of Amsterdam}
  \city{Amsterdam}
  \country{The Netherlands}
}
\email{s.s.mohammadiziabari@uva.nl}

\author{Ali Mohammed Mansoor Alsahag}
\affiliation{%
  \institution{University of Amsterdam}
  \city{Amsterdam}
  \country{The Netherlands}
}
\email{a.m.m.alsahag@uva.nl}

\begin{abstract}
Automated student-attention estimation can support learning analytics, but aggregate predictive metrics can conceal demographic disparities. This study evaluates fairness-aware multimodal temporal models on DIPSER, a naturalistic classroom dataset combining facial images, wearable-sensor measurements, attention annotations, and automatically inferred demographic metadata. Three baselines are compared across 10 training seeds: a Visual GRU, a Sensor GRU, and a Residual Fusion Transformer. The multimodal model achieves the best mean test performance (MAE 0.283, RMSE 0.363) and the lowest worst-group error among the evaluated baselines, although its gain over the Visual GRU is modest. Gender- and age-targeted MAE-gap regularization reduces disparities on validation data, but these gains do not consistently transfer to held-out subjects or repeated subject-level splits. On an NVIDIA A100-SXM4-40GB GPU, the warm end-to-end pipeline averages 50.65 ms per prediction window at a one-second stride, while the temporal model itself requires 1.02 ms. The findings show that multimodal fusion can modestly improve prediction and worst-group performance, but validation-level fairness gains should not be assumed to generalize. Robust fairness assessment therefore requires subgroup-aware evaluation, repeated subject-level validation, and larger, better balanced demographic samples.
\end{abstract}


\fancyhead{}
\maketitle

\section{Introduction}
\label{sec:introduction}

Student attention is a fundamental component of classroom learning and is strongly associated with academic performance \cite{posner_attention_2014}. In educational research, attention is commonly treated as a graded cognitive construct that reflects the extent to which students focus on learning activities \cite{olney_attention_2015} and as a core component of the broader concept of student engagement \cite{fredricks_school_2004}. As artificial intelligence becomes increasingly integrated into educational measurement and learning analytics, automated estimation of fine-grained behavioral states such as attention has become technically feasible \cite{bulut_rise_2024}. Related work in educational AI also shows that neural attention mechanisms can extract informative patterns from sequential student interactions while providing intrinsically interpretable predictions of student outcomes \cite{braakman_intrinsic_2026}. Prior work on attention estimation has shown that attentional states can be inferred from synchronized signals including facial behavior, eye tracking, physiological responses, and interaction data \cite{sharma_multimodal_2020}, with multimodal approaches often outperforming unimodal alternatives \cite{liu_individual_2025}.

Most existing student-engagement models, however, are optimized and reported primarily using aggregate predictive metrics. Such evaluation can conceal systematic differences in error across demographic groups. Bias arising from dataset composition and model design has been documented across age, gender, and race \cite{mehrabi_survey_2021}, and high-performing transformer-based models can retain substantial demographic disparities despite strong overall accuracy \cite{hosseini_faces_2025}. Fairness-aware modeling remains comparatively underexplored in student attention estimation. Thiering et al. \cite{thiering_automatic_2025}, for example, reduced gender-related distributional disparities in a vision-based engagement classifier, but their approach does not address multimodal attention estimation. This is important because physiological and behavioral modalities may provide complementary predictive information \cite{liu_individual_2025,yan_student_2025}, while multimodal fusion can itself alter subgroup performance.

This study evaluates fairness-aware multimodal temporal modeling for continuous student attention estimation on the DIPSER dataset. Three baseline architectures are compared: a visual GRU, a sensor GRU, and a Residual Fusion Transformer combining visual, physiological, and motion signals. The analysis then quantifies subgroup error disparities across inferred gender and age groups and applies an in-processing demographic MAE-gap regularizer to the strongest multimodal baseline. Because fairness constraints can modify the predictive optimum and create accuracy--fairness trade-offs \cite{menon_cost_2018}, the evaluation considers overall error, worst-group error, and best-to-worst subgroup gaps jointly rather than treating disparity reduction in isolation.

The study makes four contributions. First, it provides a controlled comparison of unimodal and multimodal temporal architectures for continuous attention estimation under identical subject-level splits and repeated random seeds. Second, it evaluates how the addition of wearable-sensor information affects both aggregate prediction error and subgroup error disparities. Third, it tests whether fairness-aware in-processing can reduce demographic error gaps without materially degrading predictive performance, including repeated subject-split analyses to assess whether validation-level fairness improvements generalize to unseen students. Finally, it benchmarks the complete inference pipeline to determine whether the proposed temporal framework can operate within the one-second prediction stride used by the dataset.

\section{Related Work}
\label{sec:related_work}

\subsection {Vision-Based Frameworks}
Research on automatic estimation of student attentional states has focused primarily on improving predictive performance using a variety of machine learning and deep learning techniques in different modalities and educational settings \cite{mandia_automatic_2024}. Early approaches were vision-based, relying primarily on facial cues, with initial studies demonstrating that engagement related states can be inferred from facial expressions \cite{whitehill_faces_2014}. Recent advances have built upon this foundation. For example, Qarbal et al. \cite{qarbal_student_2025} propose a vision-based student engagement detection framework that combines head pose estimation with facial expression analysis using transfer learning, while other CNN-based approaches trained on facial features from the DAiSEE \cite{gupta_daisee_2022} dataset incorporate dimensionality reduction, oversampling, and feature-driven model comparisons to improve engagement classification \cite{santoni_convolutional_2023} \cite{das_optimizing_2025}. Other vision-based frameworks focus on spatial facial representations by detecting faces and extracting CNN features for engagement classification in distance learning settings \cite{ferreira_development_2025}. In addition, ensemble strategies that aggregate multiple CNN models through bagging have further demonstrated an increase in performance \cite{santoni_automatic_2024}. Recent approaches extend traditional pipelines by modeling temporal facial dynamics using convolutional temporal architectures \cite{li_re-distributing_2024}, while others \cite{malekshahi_general_2024} preserve sequential frame dependencies for video-based engagement estimation, or employ transformer-based models to capture engagement dynamics across multiple temporal frequencies and address data scarcity through few-shot learning strategies \cite{mandia_transformer-driven_2025} \cite{alarefah_transformer-based_2025}. However, prior work indicates that vision-only engagement models are inherently limited, as observable facial behavior captures only a subset of students' emotional responses and offers little insight into the underlying cognitive engagement \cite{xie_multimodal_2026} \cite{noauthor_visiophysioenet_nodate}.
  
\subsection{Multimodal Fusion Frameworks}  

The inherent limitations of vision-only approaches have motivated the adoption of multimodal fusion frameworks, which aim to integrate complementary behavioral, physiological, and contextual signals through modalities such as audio, textual transcripts, wearable sensors, and interaction logs that can capture latent cognitive and affective states not directly observable from facial appearance alone. This broader motivation is also reflected in multimodal affective-computing research, where video, audio, and wearable or physiological sensing are increasingly combined to characterize affective states across real-world settings \cite{ziaee_sentiment_2026}. Prior work in education has focused on feature-level fusion of heterogeneous modalities to improve performance over single-modality models. In the broader field of student engagement classification, Yan et al. \cite{yan_student_2025} combine visual, textual, and behavioral interaction logs, resulting in a more robust multimodal pipeline. Similarly, several works integrate visual cues with physiological signals to enhance robustness. Liu et al.~\cite{liu_individual_2025} fuse CNN-based facial representations with wearable sensor data to perform continuous attention estimation in face-to-face classrooms using the DIPSER dataset, while Singh et al.~\cite{noauthor_visiophysioenet_nodate} propose VisioPhysioENet, a multimodal engagement detection framework that combines facial behavioral cues with physiological signals on the DAiSEE dataset. Both approaches demonstrate that physiological information provides complementary signals that improve performance compared to vision baselines. More advanced frameworks have explored hierarchical and temporally-aware fusion strategies to address the dynamic nature of engagement. Wang et al.~\cite{wang_classroom_2025} propose a hierarchical multimodal architecture that combines video, EEG, audio, and eye-tracking signals for classroom attention classification, with ablation studies showing that combining all modalities consistently outperforms reduced modality subsets. In a similar vein, Xie et al. propose M-LATTE \cite{xie_multimodal_2026}, a multimodal latent temporal modeling framework that performs a continuous student engagement assessment by integrating visual, audio and textual features and explicitly decomposing engagement dynamics into long-term trends and time-cyclic components using a variational autoencoder.
Furthermore, recent work has begun to explore Vision-Language and Multimodal Large Language Models as alternative multimodal integration frameworks, leveraging prompt-based inference instead of early or late fusion strategies, demonstrating that LLM-based reasoning can achieve performance comparable to classical multimodal fusion methods \cite{ma_multimodal_2025}. Using the higher-engagement experiments of the DIPSER dataset, Marquez-Carpintero et al.\cite{marquez-carpintero_enhancing_2025} report improvements over CNN baselines in attention prediction using VLMs and few-shot learning strategies.
However, recent studies indicate that Vision--Language Models are more effective on conventional emotion recognition tasks but exhibit reduced reliability when applied to engagement detection, as engagement requires contextual and commonsense reasoning beyond facial or visual cues alone \cite{teotia_evaluating_2024}. In line with this, multimodal LLMs demonstrate high accuracy in basic emotions but struggle with academic and engagement-related affect due to limited understanding of academic emotion features and insufficient ability to capture contextual information \cite{yu_exploring_2025}. Moreover, reasoning-based fusion approaches rely heavily on large pre-trained models, which have been shown to encode social and demographic biases \cite{wan_kelly_2023} \cite{wang_large_2024}, raising concerns about fairness and reliability when such models are deployed for student attention estimation.

\subsection{Fairness-Aware Learning and Evaluation}

Despite rapid progress in automatic recognition of student engagement, evaluation remains largely focused on aggregate performance metrics, with fairness-aware modeling and subgroup-level analysis remaining comparatively underexplored. This limitation is not unique to educational AI: recent transformer-based fairness analyses show that strong aggregate performance can conceal substantial label- or domain-specific disparities, motivating diagnosis-oriented evaluation beyond a single global score \cite{naranbat_fairness_2025}. Across the engagement recognition literature reviewed in this section, models are primarily evaluated using global metrics such as accuracy, weighted accuracy, F1-score, MAE, and RMSE, depending on whether engagement or attention is formulated as a classification or regression task. In their literature study on student engagement assessment, Mandia et al. \cite{mandia_automatic_2024} document extensive variation in modeling approaches and modalities, with all these approaches being consistently evaluated using global metrics without demographic disaggregation. However, recent work has highlighted that peak accuracy alone is an insufficient indicator of model reliability for the assessment of student engagement. By emphasizing predictive stability, minority-class performance, and robust interpretability, Almuniri et al. \cite{almuniri_beyond_2026} expose key limitations of single-run, accuracy-driven engagement models, motivating the need for more reliable and fairness-aware engagement estimation frameworks. In addition, Hosseini et al. \cite{hosseini_faces_2025} indicate in their findings that improvements in predictive performance in both CNN and transformer-based systems do not necessarily correspond to improvements in fairness. In particular, models that achieve state-of-the-art accuracy exhibit substantial disparities when evaluated using fairness metrics such as Equalized Odds, Equal Opportunity, Demographic Parity, and Treatment Equality \cite{hosseini_faces_2025}. Similarly, other studies demonstrate that while multimodal integration often improves predictive performance, fairness metrics may fluctuate or even deteriorate. Depending on the composition of the modalities and their availability during inference, multimodal fusion can unintentionally amplify group-level disparities rather than mitigate them \cite{sampath_multimodal_2025, schmitz_bias_2022}. As a result, explicit mitigation strategies are required to systematically identify and reduce demographic disparities. These fairness strategies in related domains can be broadly categorized into pre-processing, in-processing, and post-processing approaches.

Pre-processing approaches address bias at the data level by intervening before model training, in order to address imbalances in data distribution that could cause group disparities. Cock et al. \cite{cock_protected_2023} apply guided demographic oversampling based on combined sensitive attributes, demonstrating that intersectional balancing can mitigate bias more effectively than single-attribute correction in educational prediction tasks. Similarly, Aoudi and Al-Aqrabi (2025) \cite{aoudi_fairness-aware_2025} apply a reweighing strategy, adjusting instance weights according to the joint distribution of sensitive attributes and outcomes. While this technique preserved predictive performance, it had limited impact in some fairness evaluation metrics. This finding reflects a broader limitation of pre-processing methods, which, although computationally simple and applicable to any model architecture, may fail to fully eliminate subgroup disparities that emerge during model training.

In-processing techniques address bias directly by incorporating fairness constraints or regularization terms into the model's training objective. Related work in visual AI further demonstrates that transformer representations can encode demographic attributes such as age and sex, and that suppressing this attribute leakage must be evaluated jointly with predictive utility and worst-group performance \cite{solomon_hybrid_2026}. Kim et al. \cite{kim_fairness-aware_2023} integrate a Wasserstein distance--based regularization term to minimize distributional disparities in predicted scores across sensitive groups, complemented by adversarial representation learning to reduce sensitive attribute information in the learned embeddings. In addition, Emerson et al. \cite{emerson_multimodal_2024} employ bounded group loss constraints to enforce performance parity across demographic subgroups. Within the domain of student engagement, Thiering et al. \cite{thiering_automatic_2025} introduce attribute-orthogonal regularization in a multi-task convolutional architecture, reducing distributional disparities across gender groups in the DAiSEE video dataset, evaluated using the Pearson correlation coefficient between subgroup prediction distributions. These findings suggest that in-processing approaches, although methodologically more complex and requiring modifications to the training objective, can more directly control subgroup disparities during optimization, often achieving better results in fairness--accuracy trade-offs.

Post-processing techniques address bias by modifying the outputs of the model after training. These methods typically adjust decision thresholds or calibrate predictions to satisfy predefined fairness criteria. In multimodal settings, post-processing techniques including Equalized Odds adjustment, debiased text embeddings, and threshold optimization have demonstrated reductions in subgroup disparities, providing practical flexibility due to their applicability without model retraining, but frequently incurring documented trade-offs between fairness and predictive performance \cite{chen_exploring_2021, jiang_evaluating_2024}.

\section{Methodology}
\label{sec:methodology}

The proposed methodology is structured into four main components. The first describes the dataset and applied preprocessing steps, highlighting the multimodal structure and demographic composition relevant to fairness analysis. The second presents the design of the unimodal visual and sensor models, as well as the multimodal fusion architecture. The third describes the training procedure and evaluation metrics used for the baseline models. The fourth introduces the fairness-aware optimization framework, including the demographic disparity regularizer and the subgroup-level evaluation strategy.

\subsection{Data Preprocessing}

The data used in this work are drawn from DIPSER \cite{marquez-carpintero_dipser_2025} (Dataset for In-Person Student Engagement Recognition in the Wild), a multimodal dataset that synchronizes camera recordings with smartwatch sensor data to capture student attention and emotional states. All recordings were conducted as part of scheduled educational activities, without experimental staging or behavioral constraints, thus preserving the natural variability of student behavior. DIPSER includes both global classroom recordings and individual-level recordings, capturing facial expressions through RGB images at a resolution of 640\,$\times$\,480 pixels and a rate of approximately 10 frames per second, providing a temporal sequence of facial cues. In addition, the dataset includes wearable sensor data, comprising heart rate, linear acceleration, gyroscope, rotation vector, and ambient light measurements, capturing physiological arousal and motion dynamics at sampling rates of up to 100 Hz. The image modality is temporally aligned with per second attention labels on a scale from 1 to 5 and emotion annotations across 9 categories, obtained through a combination of expert labeling and student self-assessments. Furthermore, demographic metadata (gender, age, and ethnicity) are provided and automatically inferred using computer vision models (MiVOLO, DeepFace \cite{kuprashevich_mivolo_2023} \cite{taigman_deepface_2014}), producing probability estimates at each time frame rather than relying on self-reported information.
The dataset contains recordings from 57 subjects, organized into three groups of 16--21 students, and comprises 9 distinct educational scenarios, ranging from passive learning settings (lectures) to highly interactive activities (robotics experimentation). These sessions capture a wide range of passive and active engagement, as well as significant variation in attention, emotional states, and sensor signals. DIPSER is organized in a hierarchical structure consisting of the group, experiment, and subject levels, where the lowest level contains all modality specific data for each participant. Each subject folder includes image sequences, attention and emotion annotations (from both expert labelers and self-assessments), sensor data from wearable devices, and metadata containing demographic attributes, facial and body landmarks, and head pose estimations, all stored in JSON format. Emotion, self-reported annotations and non-demographic metadata were excluded from the analysis. To derive demographic attributes, probability estimates for gender and ethnicity were aggregated across frames. Age was calculated as the mean value and rounded to the nearest integer, and ethnicity was assigned based on the highest mean probability across all the available frames. Using this approach, the vast majority of subjects were classified as white (Caucasian), leading to the exclusion of race as a fairness axis due to this imbalance (approximately 95\% of subjects).

In addition, several preprocessing and data cleaning steps were applied prior to modeling. Invalid physiological records, such as heart rate measurements within the range of 0--30 beats per minute, were removed from the sensor streams. Subjects exhibiting a high proportion of missing sensor values within an experiment, as well as subjects without available sensor recordings, were also excluded from the final dataset. For the image modality, frames containing corrupted visual information (frames where the subject's face was not visible or outside the camera view) or lacking a corresponding sensor match were removed from the dataset prior to the modeling phase. To standardize the temporal resolution of the multimodal recordings, the first valid image frame per second was retained for each subject throughout the duration of the 5-minute experimental recordings. Sensor measurements recorded within the centered one-second interval were subsequently aggregated to produce per second sensor representations aligned with the retained visual frame. Mean values were computed for heart rate, each accelerometer and gyroscope axis, and the corresponding motion magnitudes to capture overall movement intensity. Finally, the attention label for each frame was defined as the mean score across all available expert labelers. Overall, while DIPSER provides a rich multimodal foundation for student attention estimation, its real-world setting, demographic imbalance, and the subjectivity of human annotations introduce realistic sources of bias. These characteristics make this dataset particularly suitable for investigating fairness-aware attention modeling, motivating the need for explicit bias analysis and mitigation strategies, which motivate the subgroup-aware evaluation used in this study.

\begin{table}[ht]
\centering
\caption{Demographic Distribution of Participants of the DIPSER Dataset}
\label{tab:dipser_demographics}
\small
\setlength{\tabcolsep}{4pt}
\resizebox{\columnwidth}{!}{%
\begin{tabular}{lccc}
\toprule
\textbf{Characteristic} & \textbf{Group 1 (n=20)} & \textbf{Group 2 (n=21)} & \textbf{Group 3 (n=16)} \\
\midrule

\textbf{Gender} (\%) & & & \\
Female  & 65.0 & 71.4 & 68.8 \\
Male    & 35.0 & 28.6 & 31.2 \\

\midrule
\textbf{Age (\%)} \\
13--20 & 40.0 & 52.4 & 0.0 \\
20--22 & 15.0 & 28.6 & 37.5 \\
22--26 & 35.0 & 19.0 & 25.0 \\
26--44 & 10.0 & 0.0 & 37.5 \\
\bottomrule
\end{tabular}%
}
\end{table}

\subsection{Baseline Model Architectures}

To investigate both the predictive contribution of individual modalities and their associated demographic disparities, three baseline architectures were first developed and evaluated independently: a visual model using pretrained facial image embeddings, a sensor model using physiological and motion signals from wearable devices, and a multimodal fusion model combining both modalities. All models operated on rolling temporal windows of 10 consecutive one-second observations aligned with the temporal granularity of the DIPSER annotations \cite{marquez-carpintero_dipser_2025}, allowing predictions to be updated continuously. Each window was used to estimate the level of attention in its final timestep, allowing the models to leverage the information of the preceding 10 seconds for prediction. Building on these baseline architectures, a fairness-aware multimodal transformer framework was subsequently developed for the model with the lowest mean validation error to mitigate subgroup disparities during training while preserving predictive performance.

\subsubsection{Visual Model}

A temporal visual model was developed to independently assess the predictive capability of the visual modality using temporal sequences of facial image features. The model utilized pretrained visual embeddings extracted using the CLIP-ViT-L/14 \cite{radford_learning_2021} vision encoder. Prior to feature extraction, facial images were resized to 224\,$\times$\,224 pixels using the standard CLIP preprocessing pipeline, including center cropping and normalization. For each image frame, the encoder produced a 768-dimensional visual embedding. For timestamps where no valid visual observation was available, a binary missing-frame indicator was concatenated to the visual embedding, resulting in a 769-dimensional feature vector, providing high-level semantic representations of facial appearance and behavioral cues. The resulting feature sequences were subsequently projected into a shared latent representation and processed through a recurrent temporal architecture based on a unidirectional GRU network, enabling the model to capture short-term temporal dependencies and behavioral dynamics across consecutive frames. For training stability, dropout and layer normalization were used throughout the architecture. The temporal representation generated by the GRU was subsequently passed through a regression head composed of fully connected layers with ReLU activations, producing a continuous attention score for each temporal sequence.

\subsubsection{Sensor Model}

A temporal sensor model was implemented to investigate the effectiveness of physiological and motion-based smartwatch signals in the estimation of student attention. The model utilized the wearable sensor features extracted from smartwatch recordings, including heart rate, accelerometer and gyroscope. Given the heterogeneous nature of the sensor streams, heart rate measurements and motion-related features were processed separately. Temporal sequences of motion features and heart rate values were encoded using dedicated unidirectional GRU networks to model temporal patterns within the physiological and motion signals. Invalid measurements were treated as missing observations, with binary flags used to indicate missing values during temporal processing. Consistent with the temporal visual model, dropout regularization and batch normalization were incorporated to improve generalization and reduce overfitting. The learned temporal representations from both sensor streams were subsequently fused and passed to a regression head for continuous attention estimation.

\subsubsection{Multimodal Fusion Model}

The multimodal fusion model extends the temporal visual architecture through a residual multimodal Transformer framework that integrates the physiological and motion information. The architecture uses the visual stream to produce the primary attention estimate and allows the sensor streams to provide a bounded residual correction. As in the temporal visual model, CLIP-ViT-L/14 embeddings were used to represent facial image frames, while the sensor inputs followed the same separation between motion related features and heart rate measurements adopted in the temporal sensor model. Missing or invalid sensor values were replaced with zero valued inputs after scaling, while respective binary flags were used to indicate whether the corresponding visual, motion, or heart rate observation was missing. Visual, motion, and heart rate streams were first projected into modality specific latent representations using linear projection layers. To improve training stability and generalization, the projection blocks incorporated layer normalization, GELU activations, and dropout regularization. Learnable positional embeddings were added to each modality stream to preserve temporal ordering, and dedicated causal Transformer encoders were applied independently to all modalities, enabling temporal modeling while preventing access to future observations during prediction. Following temporal encoding, the representation corresponding to the final timestep of each modality stream was retained as a summary representation of the preceding window. The visual summary representation was processed through a regression head to generate the primary attention prediction. Subsequently, the visual, motion, and heart rate summary representations were concatenated and passed to a residual fusion branch. The gating mechanism adaptively controlled the contribution of the residual correction, allowing the model to determine the extent to which sensor information should influence the final attention estimate. The final attention score was obtained by adding the gated residual correction to the visual prediction, enabling the model to preserve the robustness of the visual modality while incorporating complementary physiological and motion-based information when beneficial.

\subsection{Model Training and Evaluation}

The dataset containing 57 subjects was partitioned at the subject level to prevent data leakage, ensuring that samples from the same individual did not appear across the training, validation, and test sets. This resulted in 39 subjects for training and 9 subjects for both validation and testing. In total, the dataset comprised 285 experimental sessions and 75\,189 temporal samples for training, 64 sessions and 17\,208 samples for validation, and 59 sessions and 15\,636 samples for testing. Each sample consisted of a rolling 10-second temporal window containing visual embeddings and/or aggregated normalized sensor features, with the target defined as the attention label corresponding to the final timestep of the sequence. The subject-level split was stratified based on age group to preserve demographic representation across the different subsets. Additionally, due to the gender imbalance present in the dataset, the split was constrained to include at least 3 male subjects and all age groups in both the validation and test sets. Furthermore, due to the imbalanced distribution of attention scores within the dataset, where the majority of labels were concentrated in moderate attention ranges (scores within the range 2--3), inverse-frequency weighting was applied during training to improve the representation of underpopulated attention ranges. Attention labels were grouped into bins, and sample weights were computed inversely proportional to the frequency of each bin within the training set. To control the strength of the reweighting strategy, a weighting exponent $\alpha$ was introduced, moderating the influence of inverse-frequency weighting during optimization. The weight assigned to each attention bin $b$ was defined as:

\begin{equation}
w_b =
\left(
\frac{N}{K n_b}
\right)^\alpha,
\end{equation}

where $N$ denotes the total number of training samples, $K$ the number of attention bins, $n_b$ the number of samples within bin $b$, and $\alpha$ a hyperparameter controlling the strength of the inverse-frequency weighting.
The resulting normalized weights were incorporated into the loss function during training. The models were optimized using $\alpha=0.5$, the Adam optimizer with a learning rate of \(10^{-4}\) and weight decay of \(10^{-4}\), along with adaptive learning rate scheduling based on validation RMSE throughout the training process. Early stopping with a patience of 7 epochs was applied based on validation RMSE to reduce overfitting, and gradient clipping with a maximum norm of 1.0 was used during training to improve optimization stability.

Let \(x_i\) denote the multimodal input sample (or unimodal input in the case of the visual and sensor-only architectures), \(y_i \in [1,5]\) the ground-truth attention score, \(\hat{y}_i = f_\theta(x_i)\) the predicted attention score produced by the model with parameters \(\theta\), and \(w_i\) the inverse-frequency weight assigned to sample \(i\) according to its attention bin. The models were optimized using a weighted mean squared error objective:

\begin{equation}
\mathcal{L}_{\mathrm{task}}
=
\frac{
\sum_{i=1}^{N} w_i \left(\hat{y}_i - y_i\right)^2
}{
\sum_{i=1}^{N} w_i
},
\label{eq:task_loss}
\end{equation}

Model performance was evaluated using Mean Absolute Error (MAE) and Root Mean Squared Error (RMSE), the latter having also been used in previous works on the DIPSER dataset \cite{liu_individual_2025}. These metrics were computed both overall and between demographic subgroups. Furthermore, to assess training stability, each architecture was trained across 10 independent random seeds using the same subject-level split, and performance metrics were aggregated across runs.

\begin{table}[htbp]
\centering
\caption{Subject-level demographic distribution across the train, validation, and test splits.}
\label{tab:split_demographics}
\small
\setlength{\tabcolsep}{4pt}
\resizebox{\columnwidth}{!}{%
\begin{tabular}{lcc|cccc}
\toprule
Split & Female & Male & $(13,20]$ & $(20,22]$ & $(22,26]$ & $(26,44]$ \\
\midrule
Train & 27 & 12 & 13 & 10 & 10 & 6 \\
Validation & 6 & 3 & 3 & 3 & 2 & 1 \\
Test & 6 & 3 & 3 & 2 & 3 & 1 \\
\bottomrule
\end{tabular}%
}
\end{table}

\subsection{Deployment Feasibility Benchmark}
\label{sec:deployment_benchmark}

To evaluate whether the pipeline can support online use at the one-second prediction stride, the trained fairness-aware temporal model was benchmarked end-to-end on an unseen subject recording using an NVIDIA A100-SXM4-40GB GPU. Warm inference latency included input preprocessing, temporal image--sensor alignment, CLIP ViT-L/14 feature extraction, temporal-window construction, and the final attention prediction. Model and CLIP initialization were treated as cold-start overhead and excluded from the warm per-window latency. The temporal model was also timed separately to distinguish the cost of the learned predictor from the cost of visual feature extraction. Mean latency per prediction window and the corresponding prediction throughput were reported.

\subsection{Fairness-Aware Optimization}
\label{sec:fairness_optimization}

For the baseline architecture with the lowest mean validation error, an additional fairness-aware regularization objective was introduced to reduce predictive disparities across demographic groups. The architecture, preprocessing pipeline, subject-level split, optimizer, and evaluation setting were kept identical to the unregularized baseline model. The only modification was the introduction of a subject-level demographic fairness regularizer during training.

Let $s_i$ denote the subject identity of sample $i$, $g_s \in \mathcal{G}$ the demographic group of subject $s$, and $\mathcal{S}_g$ the set of subjects belonging to group $g$. The baseline task objective remained the weighted mean squared error loss defined in equation~\ref{eq:task_loss}. Instead of computing fairness directly at the sample level, the regularizer first computed the mean absolute error for each subject:

\begin{equation}
\text{MAE}_s
=
\frac{1}{|\mathcal{D}_s|}
\sum_{i \in \mathcal{D}_s}
\left| \hat{y}_i - y_i \right|,
\end{equation}

where $\mathcal{D}_s$ denotes the samples belonging to subject $s$. The group-level error was then obtained by averaging subject-level MAEs within each demographic group:

\begin{equation}
\text{MAE}_g
=
\frac{1}{|\mathcal{S}_g|}
\sum_{s \in \mathcal{S}_g}
\text{MAE}_s.
\end{equation}

The fairness penalty was defined as the gap between the highest and lowest group-level subject MAE:

\begin{equation}
\mathcal{L}_{\text{fair}}
=
\max_{g \in \mathcal{G}}
\text{MAE}_g
-
\min_{g \in \mathcal{G}}
\text{MAE}_g.
\end{equation}

\begin{figure}[ht]
    \centering
    \includegraphics[width=\columnwidth]{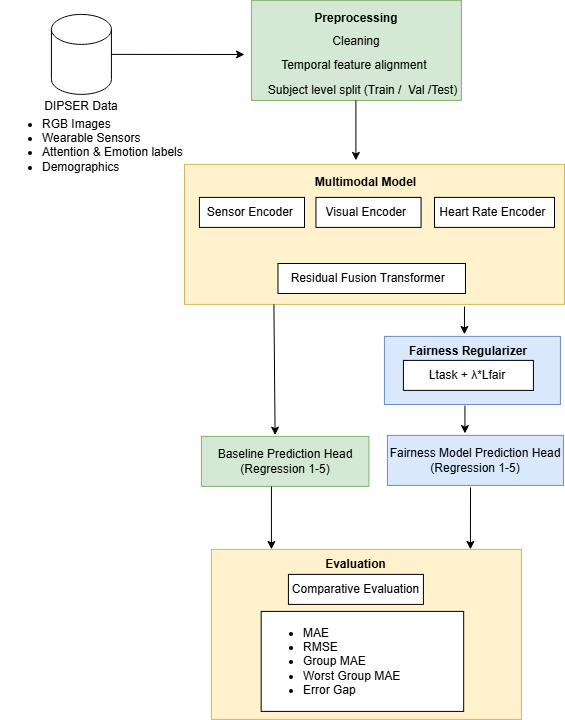}
    \Description{Pipeline showing multimodal visual and wearable-sensor inputs, temporal encoders, residual fusion, attention prediction, and demographic fairness regularization.}
    \caption{Fairness-aware multimodal attention-estimation pipeline.}
    \label{fig:methodology_framework}
\end{figure}

The final training objective was therefore:

\begin{equation}
\mathcal{L}
=
\mathcal{L}_{\text{task}}
+
\lambda \mathcal{L}_{\text{fair}},
\end{equation}

where $\lambda \geq 0$ controls the trade-off between predictive accuracy and subgroup fairness. This regularization objective penalizes unequal predictive performance between demographic subgroups, encouraging more balanced error distributions during training. To improve optimization stability, fairness regularization was introduced after a short warm-up period. During the first 2 training epochs, the model was optimized solely using the prediction MSE. The subject-level demographic MAE gap penalty was subsequently added to the objective, allowing the model to establish an initial predictive representation before balancing predictive performance and fairness. The overall pipeline of the proposed fairness-aware framework is illustrated in Figure~\ref{fig:methodology_framework}.

To obtain more stable subject-level fairness estimates, training batches were constructed using subject and group balanced sampling. For gender regularization, each batch contained 2 subjects from each gender group, with 8 temporal windows sampled from every selected subject. As a result, each gender was represented by 16 temporal windows, resulting in a total batch size of 32. For age group regularization, each batch contained 1 subject from each of the 4 age groups, again with 8 temporal windows sampled per subject, resulting in the same batch size of 32. Subjects and their corresponding temporal windows were resampled for every batch. Although the available number of temporal windows was broadly similar across subjects, this sampling procedure ensured that every represented subject contributed equally to the fairness regularizer and that the penalty was calculated using all protected groups. The regularization strength was selected using the predictions of the validation set. Candidate values of $\lambda$ were compared based on their reduction of the relevant subject-level validation MAE gap. A $\lambda$ value was considered eligible only when the mean validation MAE increased by no more than $0.02$ relative to the corresponding unregularized baseline, to ensure that fairness improvements were not achieved at the expense of substantial degradation in predictive performance. Among eligible candidates, preference was given to regularization strengths that reduced the demographic MAE gap consistently across the 10 validation seeds, followed by the magnitude of the mean gap reduction. The fairness-aware model was evaluated against the unregularized residual Transformer using overall MAE and RMSE, subgroup MAE, worst-group MAE, and the MAE gap between the best and worst performing demographic groups. Worst-group MAE was defined as the largest subgroup MAE across all demographic groups for the respective fairness axis, while MAE gap was defined as the difference between the largest and smallest subgroup MAE. These metrics quantify predictive disparities across demographic subpopulations and enable the analysis of trade-offs between overall predictive performance and subgroup fairness.

\section{Results}

\label{sec:results}

\subsection{Baseline Performance}

The predictive performance of the three architectures is summarized in Table \ref{tab:baseline_overall_performance}. The results are reported as mean and standard deviation in 10 independent training runs, using the MAE and RMSE metrics. The train mean baseline achieved an MAE of 0.312 and an RMSE of 0.393, providing a reference point for evaluating the learning-based models. The Sensor GRU achieved performance comparable to, but slightly worse than, the train mean baseline, with an MAE of 0.313 and an RMSE of 0.398. This result suggests that physiological and motion signals alone provide limited predictive information for student attention estimation in the DIPSER dataset. In contrast, the Visual GRU substantially outperformed both the train-mean baseline and the Sensor GRU, reducing the MAE to 0.287 and the RMSE to 0.366. These results suggest that facial appearance and temporal visual dynamics provide the most informative cues for the student attention prediction task. The Residual Fusion Transformer achieved the best average performance, obtaining an MAE of 0.283 and an RMSE of 0.363. However, the improvement over the Visual GRU is small and falls within the variability observed across training runs. This indicates that, although the multimodal architecture can effectively integrate sensor information, the additional physiological and motion signals provide limited complementary information beyond what is already captured by the visual modality. Consequently, visual representations appear to provide the main source of information in the attention prediction task, while sensor modalities contribute only modestly to the final performance.

\begin{table}[htbp]
\centering
\caption{Overall test performance across 10 independent training seeds.
Values are reported as mean $\pm$ standard deviation.}
\label{tab:baseline_overall_performance}
\begin{tabular}{lcc}
\toprule
Model & MAE & RMSE \\
\midrule
Train mean baseline
    & 0.312
    & 0.393 \\

Sensor GRU
    & 0.313 $\pm$ 0.002
    & 0.398 $\pm$ 0.003 \\

Visual GRU
    & 0.287 $\pm$ 0.007
    & 0.366 $\pm$ 0.008 \\

Residual Fusion Transformer
    & \textbf{0.283 $\pm$ 0.009}
    & \textbf{0.363 $\pm$ 0.009} \\
\bottomrule
\end{tabular}
\end{table}

Similarly, Table \ref{tab:baseline_subgroup_performance} presents the demographic fairness performance of the models. The Residual Fusion Transformer achieved the lowest worst-group MAE for both age (0.328) and gender (0.289), providing the most favorable balance between overall performance and worst-group error among the evaluated baselines. In addition, it exhibited the lowest gender gap (0.015), suggesting that integrating visual and sensor information through residual multimodal fusion can reduce gender disparities compared to unimodal baselines. The Sensor GRU achieved the smallest age gap (0.096), outperforming both the Visual GRU and the Residual Fusion Transformer on this fairness metric. However, this result should be interpreted together with its substantially lower predictive performance, as reported in Table \ref{tab:baseline_overall_performance}.
The relatively small age disparity may therefore partly reflect the model's overall limited predictive performance rather than an inherent improvement in fairness. In contrast, Visual GRU achieved strong overall predictive performance, but exhibited the largest age gap (0.147), suggesting that visual representations alone may be more sensitive to age related differences. This finding should also be interpreted in the context of the characteristics of DIPSER. The dataset contains 4 age groups with relatively few subjects per group, increasing the variance of subgroup-level error estimates and making both fairness evaluation and optimization more sensitive to the specific composition of the training and validation sets.

\begin{table}[htbp]
\centering
\caption{Mean demographic subgroup performance across 10 independent training seeds.
Lower values indicate better subgroup performance and smaller disparities.}
\label{tab:baseline_subgroup_performance}
\small
\resizebox{\columnwidth}{!}{
\begin{tabular}{lcccc}
\toprule
Model
& Age Worst
& Age Gap
& Gender Worst
& Gender Gap \\
\midrule
Sensor GRU
& 0.364
& \textbf{0.096}
& 0.357
& 0.068 \\

Visual GRU
& 0.341
& 0.147
& 0.295
& 0.022 \\

Residual Fusion Transformer
& \textbf{0.328}
& 0.126
& \textbf{0.289}
& \textbf{0.015} \\
\bottomrule
\end{tabular}
}
\end{table}

\subsection{Fairness-Aware Multimodal Transformer}

Based on the results of the baseline experiments, the Residual Fusion Transformer was selected for the fairness-aware framework. The fairness regularization approach described in Section~\ref{sec:fairness_optimization} was incorporated into the model's training objective, enabling the optimization process to explicitly account for demographic disparities. The following experiments evaluate the impact of this regularization on predictive performance and fairness across demographic groups. Separate models were trained for the gender and age fairness objectives, with each configuration evaluated using 10 independent training seeds. For the gender group, a regularization strength of \(\lambda=0.7\) was selected using validation performance. During validation, this configuration reduced the gender MAE gap from 0.02 to 0.005, while also slightly improving worst-group MAE, with a very small trade-off in overall predictive performance. However, these improvements did not generalize to the held-out test subjects. On the test set, both the gender MAE gap and the worst-group MAE increased compared to the unregularized model. The selected configuration reduced the gap in only 4 of 10 training runs, suggesting that these fairness improvements were not robust across random initializations and may have been sensitive to the specific composition of the training and validation data.

\begin{table}[htbp]
\centering
\caption{Validation and test performance of the unregularized and
gender MAE-gap regularized Residual Fusion Transformer across 10 independent training
seeds. Values are reported as mean $\pm$ standard deviation.}
\label{tab:gender_regularization_validation_test}
\small
\resizebox{\columnwidth}{!}{
\begin{tabular}{llccc}
\toprule
Split & Model & MAE & Gender Worst & Gender Gap \\
\midrule
Validation
& Unregularized
& \textbf{0.289 $\pm$ 0.005}
& 0.296 $\pm$ 0.009
& 0.020 $\pm$ 0.017 \\

& Gender-regularized ($\lambda=0.7$)
& 0.292 $\pm$ 0.007
& \textbf{0.294 $\pm$ 0.007}
& \textbf{0.005 $\pm$ 0.005} \\
\midrule
Test
& Unregularized
& \textbf{0.283 $\pm$ 0.009}
& \textbf{0.289 $\pm$ 0.013}
& \textbf{0.015 $\pm$ 0.016} \\

& Gender-regularized ($\lambda=0.7$)
& 0.288 $\pm$ 0.008
& 0.296 $\pm$ 0.011
& 0.022 $\pm$ 0.015 \\
\bottomrule
\end{tabular}
}
\end{table}

A similar pattern is observed for the age model, which used a regularization strength of \(\lambda=0.8\). During validation, the age-regularized model reduced the age MAE gap from 0.105 to 0.093 and improved worst-group performance from 0.352 to 0.345, while incurring only a minor increase in overall MAE. However, these improvements did not generalize to unseen data. In the test set, the regularized model exhibited worse overall predictive performance and increased demographic disparities compared to the unregularized model. Furthermore, an additional repeated subject-level split experiment was conducted to assess the robustness of the observed fairness effects across different train-validation-test splits. The dataset was partitioned multiple times at the subject level, producing different sets of training, validation, and test subjects for each run. Each split preserved the demographic structure through stratified sampling and representation constraints for the gender and age groups. The same subject partitions were then shared across the baseline and the fairness model, ensuring paired comparisons within each split. The results were consistent with the main experiments and further supported that the validation improvements achieved by fairness regularization did not correspond to consistent generalization to unseen subjects (Appendix~\ref{app:repeated-subject-splits}).

\begin{table}[htbp]
\centering
\caption{Validation and test performance of the unregularized and age MAE-gap
regularized Residual Fusion Transformer across 10 training seeds. Values are
reported as mean $\pm$ standard deviation.}
\label{tab:age_regularization_validation_test}
\small
\resizebox{\columnwidth}{!}{
\begin{tabular}{llccc}
\toprule
Split & Model & MAE & Age Worst & Age Gap \\
\midrule
Validation
& Unregularized
& \textbf{0.289 $\pm$ 0.005}
& 0.352 $\pm$ 0.018
& 0.105 $\pm$ 0.023 \\

& Age-regularized ($\lambda=0.8$)
& 0.292 $\pm$ 0.009
& \textbf{0.345 $\pm$ 0.016}
& \textbf{0.093 $\pm$ 0.013} \\
\midrule
Test
& Unregularized
& \textbf{0.283 $\pm$ 0.009}
& \textbf{0.328 $\pm$ 0.013}
& \textbf{0.126 $\pm$ 0.015} \\

& Age-regularized ($\lambda=0.8$)
& 0.292 $\pm$ 0.017
& 0.348 $\pm$ 0.020
& 0.150 $\pm$ 0.022 \\
\bottomrule
\end{tabular}
}
\end{table}

\subsection{Real-Time Feasibility}

Following the protocol in Section~\ref{sec:deployment_benchmark}, the warm end-to-end pipeline achieved a mean latency of 50.65~ms per prediction window at a one-second stride on an NVIDIA A100-SXM4-40GB GPU, corresponding to a throughput of 19.7 predictions per second. The temporal fairness model contained 402,531 trainable parameters and required 1.02~ms on average for prediction, indicating that CLIP-based visual feature extraction dominated the computational cost. These measurements show that the pipeline satisfies the one-second update requirement on the evaluated hardware. They should not, however, be interpreted as evidence of equivalent latency on lower-power or commodity deployment platforms.

\section{Discussion}
\label{sec:discussion}

\subsection{Analysis of Multimodal Attention Estimation}

The experimental results indicate that facial image representations constitute the primary source of information for student attention estimation in DIPSER \cite{marquez-carpintero_dipser_2025}. While multimodal fusion achieved the best predictive performance, the improvement over the visual GRU model was modest, with the sensor modality providing limited additional predictive information beyond the visual representation. These findings motivated the adoption of a residual fusion architecture, in which the visual stream produces the primary attention estimation, while the sensor streams contribute a bounded residual correction. In contrast, alternative fusion strategies that assigned a stronger role to the sensor modality, such as early and late fusion, did not outperform the visual GRU baseline.

In the experimental setup, the first valid image frame of each second was retained together with its corresponding sensor representation. Sensor
measurements within a centered one-second interval were aggregated by computing the mean of each sensor axis and the mean magnitudes of the accelerometer and gyroscope signals. Consequently, each 10-second sequence consisted of 10 visual frames and 10 aggregated sensor vectors. Since this design involved aggregating high-frequency wearable signals into one-second intervals, potentially removing informative short-term temporal information and thus significantly limiting the contribution of the sensor modality, an additional experiment was conducted to evaluate whether retaining all available visual frames and sensor representations would provide a measurable predictive advantage. This resulted in approximately 75 observed timesteps per 10-second sequence. Since each sensor representation summarized a centered one-second interval, sensor vectors associated with neighboring frames were highly redundant and exhibited considerable temporal overlap. This higher-frequency experiment yielded only minor improvements in the predictive evaluation metrics, while in some seeds there was no improvement (Appendix~\ref{app:fps_sensitivity}). Overall, the results provided no strong evidence that densely sampled visual frames and sensor representations offer a reliable predictive advantage over the 1 FPS setting. Instead, the findings suggest that the lower sampling frequency retains most of the information relevant for attention estimation while reducing computational cost. In addition, another important design choice was the formulation of the prediction task. Although the attention annotations are ordinal, the labels are also inherently subjective, and the boundaries between adjacent levels may not be sharply defined. For this reason, both ordinal classification and hybrid approaches were explored during preliminary experiments. Nevertheless, these alternatives did not yield improvements in the evaluation metrics and often produced less stable results across training runs.

\subsection{Demographic Disparities and Fairness Generalization}

The observed fairness patterns varied considerably between the two demographic attributes and should be interpreted in the context of the dataset composition and evaluation. More specifically, gender MAE gaps were relatively small for the Visual GRU and Residual Fusion Transformer, whereas the corresponding age gaps were substantially larger. Across both models, participants in one of the middle age groups exhibited the highest test error, while the single participant in the oldest age group had the lowest error. The observed disparity therefore does not indicate that performance deteriorated consistently with age. Instead, it reflects heterogeneous performance across the age groups represented in the test set. Part of the difference between age and gender gaps may result from the way these metrics were constructed. Age was evaluated in four groups, compared to two groups for gender. Since the disparity metric is defined as the difference between the best and worst performing groups, increasing the number of groups naturally increases the probability of observing an extreme subgroup error. 

The central finding of the fairness-aware experiments is that validation improvements in subgroup fairness did not consistently generalize to unseen subjects. During validation, the age regularizer reduced its targeted disparity across most seeds, while the gender regularizer produced more consistent improvements. However, these improvements did not transfer reliably to the test set. Although individual seeds occasionally achieved smaller test-set disparities, the targeted gaps and overall prediction errors increased on average compared to the corresponding baseline models. A possible explanation is the limited number of independent subjects available for subgroup evaluation. In particular, the test set contained only 3 male participants and a single participant in the oldest age group. Consequently, subgroup and worst-group metrics can be strongly influenced by individual participant behavior, making fairness estimates highly sensitive to the performance of a small number of subjects. The comparison between modalities provides additional context for these fairness patterns. Compared to the Visual GRU, the Residual Fusion Transformer achieved a smaller age gap, a lower worst-group error, and slightly better overall predictive performance. These results suggest that sensor information may provide complementary signals that benefit some age groups when combined with visual features, although this interpretation remains limited by the small number of subjects in several age subgroups. The Sensor GRU achieved the smallest age gap, but this came alongside the weakest overall and worst-group performance. This observation highlights that disparity metrics should be interpreted jointly with predictive performance, since lower disparities do not necessarily correspond to better outcomes for all demographic groups. Future work should evaluate fairness-aware methods using larger and more demographically balanced datasets to better assess the robustness and generalizability of fairness improvements.

\section{Conclusion}
\label{sec:conclusion}

This study evaluated whether fairness-aware multimodal temporal models can reduce demographic error disparities in student attention estimation without substantial loss of predictive performance. Across the baseline architectures, visual information provided the strongest individual modality signal. The Residual Fusion Transformer achieved the best mean test performance (MAE $0.283\pm0.009$, RMSE $0.363\pm0.009$), although its improvement over the Visual GRU was modest, indicating that the wearable-sensor streams contributed only limited complementary predictive information under the evaluated preprocessing and sampling scheme.

Subgroup evaluation revealed demographic variation that aggregate metrics alone would not expose. The Residual Fusion Transformer produced the lowest worst-group errors and the smallest gender gap among the baseline models, whereas the Sensor GRU produced the smallest age gap but substantially weaker overall performance. These results reinforce that disparity metrics should be interpreted jointly with predictive accuracy and worst-group performance rather than optimized in isolation.

Fairness regularization reduced the targeted MAE gaps on the validation set but did not consistently improve fairness for held-out subjects. The same pattern persisted across repeated subject-level splits: validation-level disparity reductions were more frequent than test-set improvements. The evidence therefore does not support a claim that the evaluated regularizer reliably generalizes demographic fairness gains to unseen students. This limitation is especially important given the small number of independent participants in several subgroups and the use of automatically inferred rather than self-reported demographic attributes.

The end-to-end pipeline also met the one-second prediction stride on an NVIDIA A100-SXM4-40GB GPU, with a mean warm latency of 50.65~ms per window; this supports real-time feasibility on the evaluated hardware but does not establish efficiency on resource-constrained devices. Overall, the results show that multimodal fusion can modestly improve attention estimation and worst-group performance, but stable fairness improvements require stronger evidence than a single validation split or a reduction in one disparity metric. Future work should prioritize larger and more demographically balanced datasets, self-reported demographic metadata where appropriate, intersectional evaluation, and alternative mitigation strategies such as reweighing \cite{aoudi_fairness-aware_2025}, adversarial debiasing, and intersectional balancing \cite{cock_protected_2023}.

\bibliographystyle{ACM-Reference-Format}
\bibliography{bibtex_acm_updated}

@article{radford_learning_2021,
	title = {Learning {Transferable} {Visual} {Models} {From} {Natural} {Language} {Supervision}},
	url = {https://www.semanticscholar.org/paper/Learning-Transferable-Visual-Models-From-Natural-Radford-Kim/6f870f7f02a8c59c3e23f407f3ef00dd1dcf8fc4},
	urldate = {2026-05-25},
	journal = {ArXiv},
	author = {Radford, Alec and Kim, Jong Wook and Hallacy, Chris and Ramesh, A. and Goh, Gabriel and Agarwal, S. and Sastry, G. and Askell, Amanda and Mishkin, Pamela and Clark, Jack and Krueger, Gretchen and Sutskever, I.},
	month = feb,
	year = {2021},
}

@misc{kuprashevich_mivolo_2023,
	title = {{MiVOLO}: {Multi}-input {Transformer} for {Age} and {Gender} {Estimation}},
	shorttitle = {{MiVOLO}},
	url = {https://arxiv.org/abs/2307.04616v2},
	language = {en},
	urldate = {2026-03-29},
	journal = {arXiv.org},
	author = {Kuprashevich, Maksim and Tolstykh, Irina},
	month = jul,
	year = {2023},
}

@techreport{olney_attention_2015,
	title = {Attention in {Educational} {Contexts}: {The} {Role} of the {Learning} {Task} in {Guiding} {Attention}},
	shorttitle = {Attention in {Educational} {Contexts}},
	url = {https://eric.ed.gov/?id=ED617714},
	language = {en},
	urldate = {2026-03-03},
	author = {Olney, Andrew M. and Risko, Evan F. and D'Mello, Sidney K. and Graesser, Arthur C.},
	year = {2015},
	note = {Publication Title: Grantee Submission
ERIC Number: ED617714},
}

@article{bulut_rise_2024,
	title = {The {Rise} of {Artificial} {Intelligence} in {Educational} {Measurement}: {Opportunities} and {Ethical} {Challenges}},
	volume = {5},
	issn = {28370899},
	shorttitle = {The {Rise} of {Artificial} {Intelligence} in {Educational} {Measurement}},
	url = {http://arxiv.org/abs/2406.18900},
	doi = {10.59863/MIQL7785},
	number = {3},
	urldate = {2026-02-22},
	journal = {Chinese/English Journal of Educational Measurement and Evaluation},
	author = {Bulut, Okan and Beiting-Parrish, Maggie and Casabianca, Jodi M. and Slater, Sharon C. and Jiao, Hong and Song, Dan and Ormerod, Christopher M. and Fabiyi, Deborah Gbemisola and Ivan, Rodica and Walsh, Cole and Rios, Oscar and Wilson, Joshua and Yildirim-Erbasli, Seyma N. and Wongvorachan, Tarid and Liu, Joyce Xinle and Tan, Bin and Morilova, Polina},
	month = dec,
	year = {2024},
	note = {arXiv:2406.18900 [cs]},
}

@article{sharma_multimodal_2020,
	title = {Multimodal data capabilities for learning: {What} can multimodal data tell us about learning?},
	volume = {51},
	copyright = {© 2020 The Authors. British Journal of Educational Technology published by John Wiley \& Sons Ltd on behalf of British Educational Research Association},
	issn = {1467-8535},
	shorttitle = {Multimodal data capabilities for learning},
	url = {https://onlinelibrary.wiley.com/doi/abs/10.1111/bjet.12993},
	doi = {10.1111/bjet.12993},
	language = {en},
	number = {5},
	urldate = {2026-02-22},
	journal = {British Journal of Educational Technology},
	author = {Sharma, Kshitij and Giannakos, Michail},
	year = {2020},
	note = {\_eprint: https://bera-journals.onlinelibrary.wiley.com/doi/pdf/10.1111/bjet.12993},
	pages = {1450--1484},
}

@article{posner_attention_2014,
	series = {3rd {Latin} {American} {Schools} on {Education} and the {Cognitive} and {Neural} {Sciences}},
	title = {Attention to learning of school subjects},
	volume = {3},
	issn = {2211-9493},
	url = {https://www.sciencedirect.com/science/article/pii/S2211949314000076},
	doi = {10.1016/j.tine.2014.02.003},
	number = {1},
	urldate = {2026-02-22},
	journal = {Trends in Neuroscience and Education},
	author = {Posner, Michael I. and Rothbart, Mary K.},
	month = mar,
	year = {2014},
	pages = {14--17},
}

@article{mehrabi_survey_2021,
	title = {A {Survey} on {Bias} and {Fairness} in {Machine} {Learning}},
	volume = {54},
	issn = {0360-0300},
	url = {https://dl.acm.org/doi/10.1145/3457607},
	doi = {10.1145/3457607},
	number = {6},
	urldate = {2026-02-21},
	journal = {ACM Comput. Surv.},
	author = {Mehrabi, Ninareh and Morstatter, Fred and Saxena, Nripsuta and Lerman, Kristina and Galstyan, Aram},
	month = jul,
	year = {2021},
	pages = {115:1--115:35},
}

@inproceedings{aoudi_fairness-aware_2025,
	title = {Fairness-{Aware} {AI} in {Education}: {Detecting} and {Reducing} {Bias} in {Student} {Assessment} {Systems}},
	shorttitle = {Fairness-{Aware} {AI} in {Education}},
	url = {https://ieeexplore.ieee.org/document/11352922},
	doi = {10.1109/ITT69610.2025.11352922},
	urldate = {2026-02-20},
	booktitle = {2025 10th {International} {Conference} on {Information} {Technology} {Trends} ({ITT})},
	author = {Aoudi, Samer and Al-Aqrabi, Hussain},
	month = nov,
	year = {2025},
	pages = {94--99},
}

@misc{thiering_automatic_2025,
	title = {Automatic {Assessment} of {Students}' {Classroom} {Engagement} with {Bias} {Mitigated} {Multi}-task {Model}},
	url = {http://arxiv.org/abs/2510.22057},
	doi = {10.48550/arXiv.2510.22057},
	urldate = {2026-02-19},
	publisher = {arXiv},
	author = {Thiering, James and Krishna, Tarun Sethupat Radha and Zelkin, Dylan and Biswas, Ashis Kumer},
	month = oct,
	year = {2025},
	note = {arXiv:2510.22057 [cs]},
}

@inproceedings{taigman_deepface_2014,
	title = {{DeepFace}: {Closing} the {Gap} to {Human}-{Level} {Performance} in {Face} {Verification}},
	issn = {1063-6919},
	shorttitle = {{DeepFace}},
	url = {https://ieeexplore.ieee.org/document/6909616},
	doi = {10.1109/CVPR.2014.220},
	urldate = {2026-02-18},
	booktitle = {2014 {IEEE} {Conference} on {Computer} {Vision} and {Pattern} {Recognition}},
	author = {Taigman, Yaniv and Yang, Ming and Ranzato, Marc'Aurelio and Wolf, Lior},
	month = jun,
	year = {2014},
	note = {ISSN: 1063-6919},
	pages = {1701--1708},
}

@article{alarefah_transformer-based_2025,
	title = {Transformer-{Based} {Student} {Engagement} {Recognition} {Using} {Few}-{Shot} {Learning}},
	volume = {14},
	copyright = {http://creativecommons.org/licenses/by/3.0/},
	issn = {2073-431X},
	url = {https://www.mdpi.com/2073-431X/14/3/109},
	doi = {10.3390/computers14030109},
	language = {en},
	number = {3},
	urldate = {2026-02-16},
	journal = {Computers},
	publisher = {publisher},
	author = {Alarefah, Wejdan and Jarraya, Salma Kammoun and Abuzinadah, Nihal},
	month = mar,
	year = {2025},
}

@inproceedings{menon_cost_2018,
	title = {The cost of fairness in binary classification},
	issn = {2640-3498},
	url = {https://proceedings.mlr.press/v81/menon18a.html},
	language = {en},
	urldate = {2026-02-16},
	booktitle = {Proceedings of the 1st {Conference} on {Fairness}, {Accountability} and {Transparency}},
	publisher = {PMLR},
	author = {Menon, Aditya Krishna and Williamson, Robert C.},
	month = jan,
	year = {2018},
	pages = {107--118},
}

@article{fredricks_school_2004,
	title = {School {Engagement}: {Potential} of the {Concept}, {State} of the {Evidence}},
	volume = {74},
	issn = {0034-6543},
	shorttitle = {School {Engagement}},
	url = {https://www.jstor.org/stable/3516061},
	number = {1},
	urldate = {2026-02-16},
	journal = {Review of Educational Research},
	publisher = {[Sage Publications, Inc., American Educational Research Association]},
	author = {Fredricks, Jennifer A. and Blumenfeld, Phyllis C. and Paris, Alison H.},
	year = {2004},
	pages = {59--109},
}

@techreport{emerson_multimodal_2024,
	title = {Multimodal, {Multi}-{Class} {Bias} {Mitigation} for {Predicting} {Speaker} {Confidence}},
	url = {https://eric.ed.gov/?id=ED675542},
	language = {en},
	urldate = {2026-02-16},
	institution = {International Educational Data Mining Society},
	author = {Emerson, Andrew and Ramesh, Arti and Houghton, Patrick and Basheerabad, Vinay and Jawahar, Navaneeth and Leong, Chee Wee},
	year = {2024},
	note = {ERIC Number: ED675542},
}

@article{kim_fairness-aware_2023,
	title = {Fairness-{Aware} {Multimodal} {Learning} in {Automatic} {Video} {Interview} {Assessment}},
	volume = {11},
	issn = {2169-3536},
	url = {https://ieeexplore.ieee.org/document/10287972},
	doi = {10.1109/ACCESS.2023.3325891},
	urldate = {2026-02-16},
	journal = {IEEE Access},
	author = {Kim, Changwoo and Choi, Jinho and Yoon, Jongyeon and Yoo, Daehun and Lee, Woojin},
	year = {2023},
	pages = {122677--122693},
}

@misc{hosseini_faces_2025,
	title = {Faces of {Fairness}: {Examining} {Bias} in {Facial} {Expression} {Recognition} {Datasets} and {Models}},
	shorttitle = {Faces of {Fairness}},
	url = {http://arxiv.org/abs/2502.11049},
	doi = {10.48550/arXiv.2502.11049},
	urldate = {2026-02-16},
	publisher = {arXiv},
	author = {Hosseini, Mohammad Mehdi and Fard, Ali Pourramezan and Mahoor, Mohammad H.},
	month = oct,
	year = {2025},
	note = {arXiv:2502.11049 [cs]},
}

@inproceedings{cock_protected_2023,
	address = {New York, NY, USA},
	series = {{LAK2023}},
	title = {Protected {Attributes} {Tell} {Us} {Who}, {Behavior} {Tells} {Us} {How}: {A} {Comparison} of {Demographic} and {Behavioral} {Oversampling} for {Fair} {Student} {Success} {Modeling}},
	isbn = {978-1-4503-9865-7},
	shorttitle = {Protected {Attributes} {Tell} {Us} {Who}, {Behavior} {Tells} {Us} {How}},
	url = {https://dl.acm.org/doi/10.1145/3576050.3576149},
	doi = {10.1145/3576050.3576149},
	urldate = {2026-02-15},
	booktitle = {{LAK23}: 13th {International} {Learning} {Analytics} and {Knowledge} {Conference}},
	publisher = {Association for Computing Machinery},
	author = {Cock, Jade Mai and Bilal, Muhammad and Davis, Richard and Marras, Mirko and Kaser, Tanja},
	month = mar,
	year = {2023},
	pages = {488--498},
}

@article{yu_exploring_2025,
	title = {Exploring the prospects of multimodal large language models for {Automated} {Emotion} {Recognition} in education: {Insights} from {Gemini}},
	volume = {232},
	issn = {0360-1315},
	shorttitle = {Exploring the prospects of multimodal large language models for {Automated} {Emotion} {Recognition} in education},
	url = {https://www.sciencedirect.com/science/article/pii/S0360131525000752},
	doi = {10.1016/j.compedu.2025.105307},
	urldate = {2026-02-12},
	journal = {Computers \& Education},
	author = {Yu, Shuzhen and Androsov, Alexey and Yan, Hanbing},
	month = jul,
	year = {2025},
	pages = {105307},
}

@misc{schmitz_bias_2022,
	title = {Bias and {Fairness} on {Multimodal} {Emotion} {Detection} {Algorithms}},
	url = {http://arxiv.org/abs/2205.08383},
	doi = {10.13140/RG.2.2.14341.01769},
	urldate = {2026-02-12},
	author = {Schmitz, Matheus and Ahmed, Rehan and Cao, Jimi},
	month = may,
	year = {2022},
	note = {arXiv:2205.08383 [cs]},
}

@article{jiang_evaluating_2024,
	title = {Evaluating and mitigating unfairness in multimodal remote mental health assessments},
	volume = {3},
	issn = {2767-3170},
	url = {https://journals.plos.org/digitalhealth/article?id=10.1371/journal.pdig.0000413},
	doi = {10.1371/journal.pdig.0000413},
	language = {en},
	number = {7},
	urldate = {2026-02-12},
	journal = {PLOS Digital Health},
	publisher = {Public Library of Science},
	author = {Jiang, Zifan and Seyedi, Salman and Griner, Emily and Abbasi, Ahmed and Rad, Ali Bahrami and Kwon, Hyeokhyen and Cotes, Robert O. and Clifford, Gari D.},
	month = jul,
	year = {2024},
	pages = {e0000413},
}

@misc{chen_exploring_2021,
	title = {Exploring {Text} {Specific} and {Blackbox} {Fairness} {Algorithms} in {Multimodal} {Clinical} {NLP}},
	url = {http://arxiv.org/abs/2011.09625},
	doi = {10.48550/arXiv.2011.09625},
	urldate = {2026-02-12},
	publisher = {arXiv},
	author = {Chen, John and Berlot-Attwell, Ian and Hossain, Safwan and Wang, Xindi and Rudzicz, Frank},
	month = jun,
	year = {2021},
	note = {arXiv:2011.09625 [cs]},
}

@misc{sampath_multimodal_2025,
	title = {The {Multimodal} {Paradox}: {How} {Added} and {Missing} {Modalities} {Shape} {Bias} and {Performance} in {Multimodal} {AI}},
	shorttitle = {The {Multimodal} {Paradox}},
	url = {http://arxiv.org/abs/2505.03020},
	doi = {10.48550/arXiv.2505.03020},
	urldate = {2026-02-11},
	publisher = {arXiv},
	author = {Sampath, Kishore and Pratheesh and Mohammad, Ayaazuddin and Ramachandranpillai, Resmi},
	month = may,
	year = {2025},
	note = {arXiv:2505.03020 [cs]},
}

@article{mandia_automatic_2024,
	title = {Automatic student engagement measurement using machine learning techniques: {A} literature study of data and methods},
	volume = {83},
	issn = {1573-7721},
	shorttitle = {Automatic student engagement measurement using machine learning techniques},
	url = {https://doi.org/10.1007/s11042-023-17534-9},
	doi = {10.1007/s11042-023-17534-9},
	language = {en},
	number = {16},
	urldate = {2026-02-11},
	journal = {Multimedia Tools and Applications},
	author = {Mandia, Sandeep and Mitharwal, Rajendra and Singh, Kuldeep},
	month = may,
	year = {2024},
	pages = {49641--49672},
}

@inproceedings{teotia_evaluating_2024,
	title = {Evaluating {Vision} {Language} {Models} in {Detecting} {Learning} {Engagement}},
	issn = {2375-9259},
	url = {https://ieeexplore.ieee.org/abstract/document/10917661},
	doi = {10.1109/ICDMW65004.2024.00069},
	urldate = {2026-02-09},
	booktitle = {2024 {IEEE} {International} {Conference} on {Data} {Mining} {Workshops} ({ICDMW})},
	author = {Teotia, Jayant and Zhang, Xulang and Mao, Rui and Cambria, Erik},
	month = dec,
	year = {2024},
	note = {ISSN: 2375-9259},
	pages = {496--502},
}

@article{ferreira_development_2025,
	title = {Development of a framework using deep learning for the identification and classification of engagement levels in distance learning students},
	volume = {15},
	issn = {1869-5469},
	url = {https://doi.org/10.1007/s13278-025-01408-z},
	doi = {10.1007/s13278-025-01408-z},
	language = {en},
	number = {1},
	urldate = {2026-02-09},
	journal = {Social Network Analysis and Mining},
	author = {Ferreira, Fernando Rodrigues Trindade and do Couto, Loena Marins and de Melo Baptista Domingues, Guilherme and Saporetti, Camila Martins},
	month = apr,
	year = {2025},
	pages = {37},
}

@misc{mandia_transformer-driven_2025,
	title = {Transformer-{Driven} {Modeling} of {Variable} {Frequency} {Features} for {Classifying} {Student} {Engagement} in {Online} {Learning}},
	url = {http://arxiv.org/abs/2502.10813},
	doi = {10.48550/arXiv.2502.10813},
	urldate = {2026-02-09},
	publisher = {arXiv},
	author = {Mandia, Sandeep and Singh, Kuldeep and Mitharwal, Rajendra and Mushtaq, Faisel and Janu, Dimpal},
	month = feb,
	year = {2025},
	note = {arXiv:2502.10813 [cs]},
}

@misc{malekshahi_general_2024,
	title = {A {General} {Model} for {Detecting} {Learner} {Engagement}: {Implementation} and {Evaluation}},
	shorttitle = {A {General} {Model} for {Detecting} {Learner} {Engagement}},
	url = {http://arxiv.org/abs/2405.04251},
	doi = {10.48550/arXiv.2405.04251},
	urldate = {2026-02-08},
	publisher = {arXiv},
	author = {Malekshahi, Somayeh and Kheyridoost, Javad M. and Fatemi, Omid},
	month = may,
	year = {2024},
	note = {arXiv:2405.04251 [cs]},
}

@inproceedings{wan_kelly_2023,
	address = {Singapore},
	title = {“{Kelly} is a {Warm} {Person}, {Joseph} is a {Role} {Model}”: {Gender} {Biases} in {LLM}-{Generated} {Reference} {Letters}},
	shorttitle = {“{Kelly} is a {Warm} {Person}, {Joseph} is a {Role} {Model}”},
	url = {https://aclanthology.org/2023.findings-emnlp.243/},
	doi = {10.18653/v1/2023.findings-emnlp.243},
	urldate = {2026-02-08},
	booktitle = {Findings of the {Association} for {Computational} {Linguistics}: {EMNLP} 2023},
	publisher = {Association for Computational Linguistics},
	author = {Wan, Yixin and Pu, George and Sun, Jiao and Garimella, Aparna and Chang, Kai-Wei and Peng, Nanyun},
	editor = {Bouamor, Houda and Pino, Juan and Bali, Kalika},
	month = dec,
	year = {2023},
	pages = {3730--3748},
}

@inproceedings{wang_large_2024,
	address = {Bangkok, Thailand},
	title = {Large {Language} {Models} are not {Fair} {Evaluators}},
	url = {https://aclanthology.org/2024.acl-long.511/},
	doi = {10.18653/v1/2024.acl-long.511},
	urldate = {2026-02-08},
	booktitle = {Proceedings of the 62nd {Annual} {Meeting} of the {Association} for {Computational} {Linguistics} ({Volume} 1: {Long} {Papers})},
	publisher = {Association for Computational Linguistics},
	author = {Wang, Peiyi and Li, Lei and Chen, Liang and Cai, Zefan and Zhu, Dawei and Lin, Binghuai and Cao, Yunbo and Kong, Lingpeng and Liu, Qi and Liu, Tianyu and Sui, Zhifang},
	editor = {Ku, Lun-Wei and Martins, Andre and Srikumar, Vivek},
	month = aug,
	year = {2024},
	pages = {9440--9450},
}

@inproceedings{ma_multimodal_2025,
	title = {Multimodal {Fusion} with {LLMs} for {Engagement} {Prediction} in {Natural} {Conversation}},
	url = {http://arxiv.org/abs/2409.09135},
	doi = {10.1145/3747327.3764904},
	urldate = {2026-02-08},
	booktitle = {Companion {Proceedings} of the 27th {International} {Conference} on {Multimodal} {Interaction}},
	author = {Ma, Cheng Charles and Joo, Kevin Hyekang and Vail, Alexandria K. and Bhattacharya, Sunreeta and García, Álvaro Fernández and Baker-Matsuoka, Kailana and Mathew, Sheryl and Holt, Lori L. and Torre, Fernando De la},
	month = oct,
	year = {2025},
	note = {arXiv:2409.09135 [cs]},
	pages = {244--259},
}

@article{noauthor_visiophysioenet_nodate,
  author = {Singh, Alakhsimar and Verma, Nischay and Goyal, Kanav and Singh, Amritpal and Kumar, Puneet and Li, Xiaobai},
  title = {VisioPhysioENet: Multimodal Engagement Detection using Visual and Physiological Signals},
  journal = {arXiv preprint arXiv:2409.16126},
  year = {2024},
  doi = {10.48550/arXiv.2409.16126},
  url = {https://arxiv.org/abs/2409.16126}
}

@article{whitehill_faces_2014,
	title = {The {Faces} of {Engagement}: {Automatic} {Recognition} of {Student} {Engagementfrom} {Facial} {Expressions}},
	volume = {5},
	issn = {1949-3045},
	shorttitle = {The {Faces} of {Engagement}},
	url = {https://ieeexplore.ieee.org/document/6786307},
	doi = {10.1109/TAFFC.2014.2316163},
	number = {1},
	urldate = {2026-02-07},
	journal = {IEEE Transactions on Affective Computing},
	author = {Whitehill, Jacob and Serpell, Zewelanji and Lin, Yi-Ching and Foster, Aysha and Movellan, Javier R.},
	month = jan,
	year = {2014},
	pages = {86--98},
}

@article{almuniri_beyond_2026,
	title = {Beyond peak accuracy: a stability-centric framework for reliable multimodal student engagement assessment},
	volume = {16},
	copyright = {2026 The Author(s)},
	issn = {2045-2322},
	shorttitle = {Beyond peak accuracy},
	url = {https://www.nature.com/articles/s41598-025-31215-7},
	doi = {10.1038/s41598-025-31215-7},
	language = {en},
	number = {1},
	urldate = {2026-02-06},
	journal = {Scientific Reports},
	publisher = {Nature Publishing Group},
	author = {Almuniri, Ismail Said and Alhussian, Hitham and Aziz, Norshakirah and Khairy, Sallam O. F. and AlAbri, AlWaleed Sulaiman and Jarallah, Zaid Fawaz and Yahaya, Saidu and Adamu, Shamsuddeen},
	month = jan,
	year = {2026},
	pages = {5},
}

@article{li_re-distributing_2024,
	title = {Re-{Distributing} {Facial} {Features} for {Engagement} {Prediction} with {ModernTCN}},
	volume = {81},
	issn = {1546-2218, 1546-2226},
	url = {https://www.techscience.com/cmc/v81n1/58326},
	doi = {10.32604/cmc.2024.054982},
	language = {en},
	number = {1},
	urldate = {2026-02-06},
	journal = {Computers, Materials \& Continua},
	publisher = {Tech Science Press},
	author = {Li, Xi and Zhu, Weiwei and Li, Qian and Hou, Changhui and Zhang, Yaozong},
	year = {2024},
	pages = {369--391},
}

@article{yan_student_2025,
	title = {Student engagement assessment using multimodal deep learning},
	volume = {20},
	issn = {1932-6203},
	url = {https://journals.plos.org/plosone/article?id=10.1371/journal.pone.0325377},
	doi = {10.1371/journal.pone.0325377},
	language = {en},
	number = {6},
	urldate = {2026-02-06},
	journal = {PLOS ONE},
	publisher = {Public Library of Science},
	author = {Yan, Lijuan and Wu, Xiaotao and Wang, Yi},
	month = jun,
	year = {2025},
	pages = {e0325377},
}

@inproceedings{qarbal_student_2025,
	address = {Cham},
	title = {Student {Engagement} {Detection} {Based} on {Head} {Pose} {Estimation} and {Facial} {Expressions} {Using} {Transfer} {Learning}},
	isbn = {978-3-031-88653-9},
	doi = {10.1007/978-3-031-88653-9_25},
	language = {en},
	booktitle = {Innovations in {Smart} {Cities} {Applications} {Volume} 8},
	publisher = {Springer Nature Switzerland},
	author = {Qarbal, Ikram and Sael, Nawal and Ouahabi, Sara},
	editor = {Ben Ahmed, Mohamed and Abdelhakim, Boudhir Anouar and Karaș, İsmail Rakıp and Ben Ahmed, Kaoutar},
	year = {2025},
	pages = {246--255},
}

@article{das_optimizing_2025,
	title = {Optimizing student engagement detection using facial and behavioral features},
	volume = {37},
	issn = {1433-3058},
	url = {https://doi.org/10.1007/s00521-025-11317-z},
	doi = {10.1007/s00521-025-11317-z},
	language = {en},
	number = {23},
	urldate = {2026-02-05},
	journal = {Neural Computing and Applications},
	author = {Das, Riju and Dev, Soumyabrata},
	month = aug,
	year = {2025},
	pages = {19063--19085},
}

@inproceedings{wang_classroom_2025,
	title = {Classroom {Attention} {Detection} {Model} through {Multimodal} {Data} {Fusion}},
	url = {https://ieeexplore.ieee.org/document/11329999},
	doi = {10.1109/DSInS68311.2025.11329999},
	urldate = {2026-02-05},
	booktitle = {2025 5th {International} {Conference} on {Digital} {Society} and {Intelligent} {Systems} ({DSInS})},
	author = {Wang, Wanjun and Yan, Yan and Ding, Runxia and Li, Yuan and Zhang, Hui and Du, Lei},
	month = nov,
	year = {2025},
	pages = {310--314},
}

@article{santoni_automatic_2024,
	title = {Automatic {Detection} of {Students}’ {Engagement} {During} {Online} {Learning}: {A} {Bagging} {Ensemble} {Deep} {Learning} {Approach}},
	volume = {12},
	issn = {2169-3536},
	shorttitle = {Automatic {Detection} of {Students}’ {Engagement} {During} {Online} {Learning}},
	url = {https://ieeexplore.ieee.org/document/10591967},
	doi = {10.1109/ACCESS.2024.3425820},
	urldate = {2026-02-05},
	journal = {IEEE Access},
	author = {Santoni, Mayanda Mega and Basaruddin, T. and Junus, Kasiyah and Lawanto, Oenardi},
	year = {2024},
	pages = {96063--96073},
}

@misc{gupta_daisee_2022,
	title = {{DAiSEE}: {Towards} {User} {Engagement} {Recognition} in the {Wild}},
	shorttitle = {{DAiSEE}},
	url = {http://arxiv.org/abs/1609.01885},
	doi = {10.48550/arXiv.1609.01885},
	urldate = {2026-02-05},
	publisher = {arXiv},
	author = {Gupta, Abhay and D'Cunha, Arjun and Awasthi, Kamal and Balasubramanian, Vineeth},
	month = jul,
	year = {2022},
	note = {arXiv:1609.01885 [cs]},
}

@article{santoni_convolutional_2023,
	title = {Convolutional {Neural} {Network} {Model} based {Students}’ {Engagement} {Detection} in {Imbalanced} {DAiSEE} {Dataset}},
	volume = {14},
	issn = {2156-5570},
	url = {https://thesai.org/Publications/ViewPaper?Volume=14&Issue=3&Code=IJACSA&SerialNo=71},
	doi = {10.14569/IJACSA.2023.0140371},
	language = {en},
	number = {3},
	urldate = {2026-02-05},
	journal = {International Journal of Advanced Computer Science and Applications (IJACSA)},
	publisher = {The Science and Information (SAI) Organization Limited},
	author = {Santoni, Mayanda Mega and Basaruddin, T. and Junus, Kasiyah},
	month = mar,
	year = {2023},
}

@article{xie_multimodal_2026,
	title = {Multimodal {Latent} {Temporal} {Modeling} for {Continuous} {Engagement} {Assessment} in {Online} {Education}},
	issn = {1939-1382},
	url = {https://ieeexplore.ieee.org/abstract/document/11359967},
	doi = {10.1109/TLT.2026.3656606},
	urldate = {2026-02-05},
	journal = {IEEE Transactions on Learning Technologies},
	author = {Xie, Congcong and Wang, Di and Wang, Quan and Liang, Xiao and Liu, Ruyi and Miao, Qiguang},
	year = {2026},
	pages = {1--14},
}

@article{marquez-carpintero_enhancing_2025,
	title = {Enhancing {Engineering} and {STEM} {Education} {With} {Vision} and {Multimodal} {Large} {Language} {Models} to {Predict} {Student} {Attention}},
	volume = {13},
	issn = {2169-3536},
	url = {https://ieeexplore.ieee.org/document/11053810},
	doi = {10.1109/ACCESS.2025.3584025},
	urldate = {2026-02-04},
	journal = {IEEE Access},
	author = {Marquez-Carpintero, Luis and Viejo, Diego and Cazorla, Miguel},
	year = {2025},
	pages = {114681--114695},
}

@inproceedings{liu_individual_2025,
	title = {Individual {Student} {Attention} {Detection} in {Face}-to-{Face} {Classrooms} {Using} {Multimodal} {Facial} and {Wearable} {Data}},
	issn = {2766-2144},
	url = {https://ieeexplore.ieee.org/abstract/document/11113529},
	doi = {10.1109/ISET65607.2025.00017},
	urldate = {2026-02-02},
	booktitle = {2025 {International} {Symposium} on {Educational} {Technology} ({ISET})},
	author = {Liu, Jiaqi and Chui, Kwok Tai and Lee, Lap–Kei and Ng, Kwan Keung and Paoprasert, Naraphorn and Zhao, Mingbo},
	month = jul,
	year = {2025},
	note = {ISSN: 2766-2144},
	pages = {38--43},
}

@misc{marquez-carpintero_dipser_2025,
	title = {{DIPSER}: {A} {Dataset} for {In}-{Person} {Student} {Engagement} {Recognition} in the {Wild}},
	shorttitle = {{DIPSER}},
	url = {http://arxiv.org/abs/2502.20209},
	doi = {10.48550/arXiv.2502.20209},
	urldate = {2026-01-31},
	publisher = {arXiv},
	author = {Marquez-Carpintero, Luis and Suescun-Ferrandiz, Sergio and Álvarez, Carolina Lorenzo and Fernandez-Herrero, Jorge and Viejo, Diego and Roig-Vila, Rosabel and Cazorla, Miguel},
	month = mar,
	year = {2025},
	note = {arXiv:2502.20209 [cs]},
}

@article{braakman_intrinsic_2026,
  author  = {Braakman, Nina May and Mohammadi Ziabari, Seyed Sahand and Alsahag, Ali Mohammed Mansoor and Al Husaini, Yousuf Nasser},
  title   = {Intrinsic Interpretability at Parity: Attention-Based {RL--MIL} for Student Outcome Prediction},
  journal = {Natural Language Processing Journal},
  volume  = {14},
  pages   = {100204},
  year    = {2026},
  doi     = {10.1016/j.nlp.2026.100204},
  url     = {https://doi.org/10.1016/j.nlp.2026.100204}
}

@article{ziaee_sentiment_2026,
  author  = {Ziaee, Ali and Mohammadi Ziabari, Seyed Sahand},
  title   = {Sentiment Analysis of Sports Fans' Behaviour: A Multimodal Review of Digital and In-Venue Fan Affect},
  journal = {Journal of Sports Analytics},
  volume  = {12},
  year    = {2026},
  doi     = {10.1177/22150218261455495},
  url     = {https://doi.org/10.1177/22150218261455495}
}

@misc{naranbat_fairness_2025,
  author        = {Naranbat, Battemuulen and Mohammadi Ziabari, Seyed Sahand and Al Husaini, Yousuf Nasser and Alsahag, Ali Mohammed Mansoor},
  title         = {Fairness Metric Design Exploration in Multi-Domain Moral Sentiment Classification using Transformer-Based Models},
  year          = {2025},
  eprint        = {2510.11222},
  archivePrefix = {arXiv},
  primaryClass  = {cs.CL},
  doi           = {10.48550/arXiv.2510.11222},
  url           = {https://arxiv.org/abs/2510.11222}
}

@misc{solomon_hybrid_2026,
  author        = {Solomon, Jobeal and Alsahag, Ali Mohammed Mansoor and Mohammadi Ziabari, Seyed Sahand},
  title         = {Hybrid Vision Transformer--{GAN} Attribute Neutralizer for Mitigating Bias in Chest {X}-Ray Diagnosis},
  year          = {2026},
  eprint        = {2601.15490},
  archivePrefix = {arXiv},
  primaryClass  = {cs.CV},
  doi           = {10.48550/arXiv.2601.15490},
  url           = {https://arxiv.org/abs/2601.15490}
}

\clearpage
\appendix
\onecolumn

\section{Training and Validation Error Curves}
\label{sec:apx:first_appendix}
\label{app:learning_curves}

\begin{figure}[htbp]
    \centering
    \includegraphics[width=\textwidth]
    {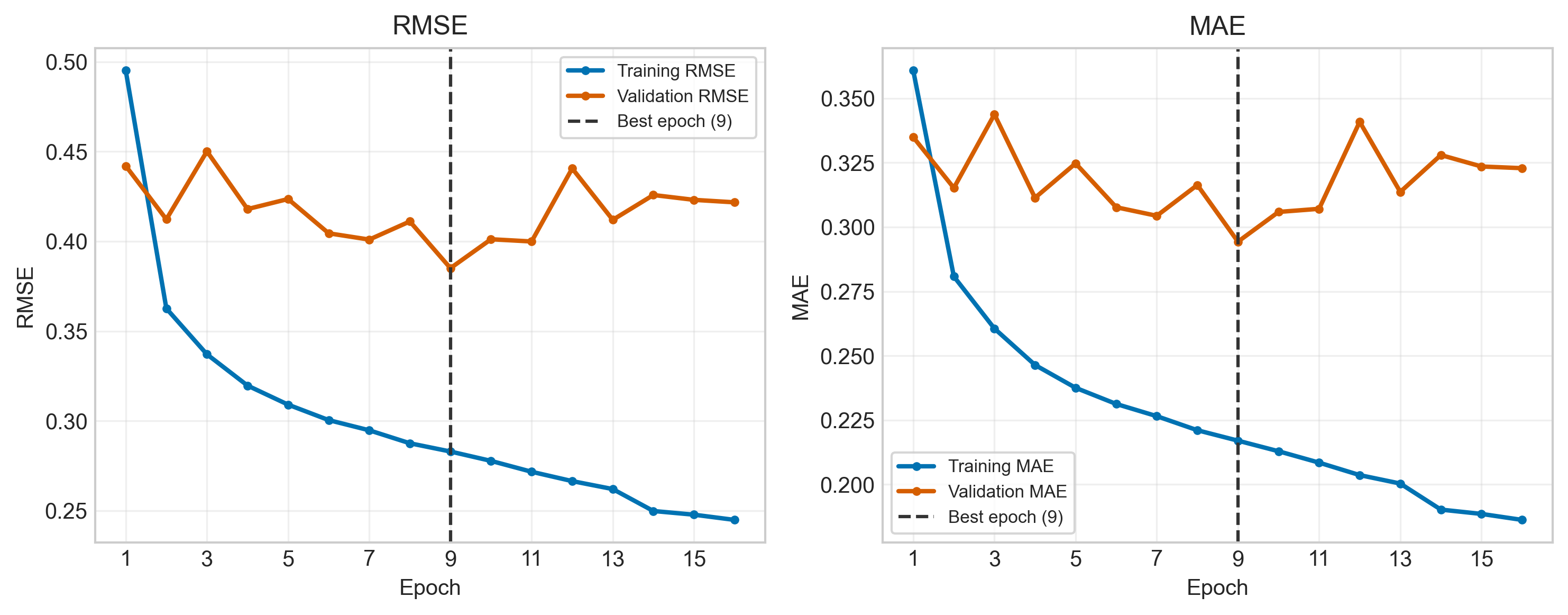}
    \Description{Training and validation MAE and RMSE curves over epochs for the Residual Fusion Transformer.}
    \caption{Training and validation MAE and RMSE curves for the
    Residual Fusion Transformer using seed 100. The dashed line
    indicates the epoch selected based on the lowest validation RMSE.}
    \label{fig:residual_learning_curves}
\end{figure}

Figure~\ref{fig:residual_learning_curves} presents the training and
validation MAE and RMSE curves for a representative Residual Fusion
Transformer run. Seed 100 was selected because its best validation
RMSE was closest to the median result across the 10 independent
runs. The increasing separation between the training and validation
curves after epoch 9 supports the use of
validation-based early stopping.

\section{Repeated Subject-Split Robustness Analysis}
\label{app:repeated-subject-splits}

\begin{table}[htbp]
\centering
\small
\caption{Repeated subject-split fairness robustness analysis. Values are paired changes relative to the unregularized Residual Fusion Transformer across 10 repeated subject-level splits. Positive values indicate improvement over the baseline.}
\label{tab:repeated-split-fairness}
\begin{tabular}{llcc}
\toprule
Fairness target & Metric & Mean change vs. baseline & Positive splits \\
\midrule
Age & Validation overall MAE gain & -0.003 & 3/10 \\
Age & Validation age gap reduction & 0.029 & 9/10 \\

Age & Test overall MAE gain & 0.006 & 6/10 \\
Age & Test age gap reduction & -0.007 & 2/10 \\

\midrule
Gender & Validation overall MAE gain & -0.004 & 3/10 \\
Gender & Validation gender gap reduction & 0.010 & 6/10 \\
Gender & Test overall MAE gain & 0.003 & 6/10 \\
Gender & Test gender gap reduction & -0.012 & 3/10 \\

\bottomrule
\end{tabular}
\end{table}

\section{Frame Sampling Rate Sensitivity Analysis}
\label{app:fps_sensitivity}

\begin{table}[ht]
\centering
\caption{Comparison of 1 frame per second and 10 frames per second using the Residual Transformer across 5 matched seeds. Values are reported as means across the matched seeds.}
\label{tab:fps_comparison}
\begin{tabular}{lccc}
\hline
Data & 1 FPS & 10 FPS  \\
\hline
Validation MAE  & 0.289 & 0.291 \\
Validation RMSE & 0.385 & 0.386 \\
Test MAE        & 0.283 & 0.282 \\
Test RMSE       & 0.363 & 0.361 \\
\hline
\end{tabular}
\end{table}

\clearpage
\section{Architecture and Training Hyperparameters}
\label{app:architecture-hyperparameters}

\begin{table}[H]
\centering
\small
\caption{Architecture hyperparameters of the baseline and multimodal models. The fairness-aware models use the same Residual Fusion Transformer architecture as the unregularized multimodal model; only the training objective is modified by the demographic MAE-gap regularizer.}
\label{tab:architecture-hyperparameters}
\begin{tabular}{p{3.0cm}p{4.0cm}p{4.0cm}p{5.2cm}}
\toprule
Component & Visual GRU & Sensor GRU & Residual Fusion Transformer \\
\midrule
Input representation
& CLIP ViT-L/14 embedding: $768$ dimensions + visual missing flag
& Motion stream: motion features + missing flag; heart-rate stream: heart rate + missing flag
& Visual stream: $768$-dimensional CLIP embedding + visual missing flag; separate motion and heart-rate streams with missingness flags \\

Projection layers
& Linear $769 \rightarrow 512$, LayerNorm, ReLU, dropout $0.2$
& No projection before GRU streams
& Visual projection $769 \rightarrow 128$; motion projection $\rightarrow 32$; heart-rate projection $\rightarrow 16$; LayerNorm and GELU used in all streams \\

Temporal encoder
& One-layer GRU
& Separate one-layer GRUs for motion and heart rate
& Separate causal Transformer encoders for visual, motion, and heart-rate streams \\

Hidden dimensions
& GRU hidden size $128$
& Motion GRU hidden size $64$; heart-rate GRU hidden size $16$
& Visual dimension $128$; motion dimension $32$; heart-rate dimension $16$ \\

Transformer layers
& - 
& - 
& Visual: $2$ layers; motion: $1$ layer; heart rate: $1$ layer \\

Attention heads
& - 
& - 
& Visual: $4$ heads; motion: $4$ heads; heart rate: $2$ heads \\

Feed-forward dimensions
& - 
& - 
& Visual: $256$; motion: $64$; heart rate: $32$ \\

Prediction head
& Linear $128 \rightarrow 64 \rightarrow 1$
& Linear $80 \rightarrow 64 \rightarrow 32 \rightarrow 1$
& Visual prediction head $128 \rightarrow 64 \rightarrow 1$; residual correction head $176 \rightarrow 64 \rightarrow 1$ \\

Residual gating
& - 
& - 
& Gate network $176 \rightarrow 32 \rightarrow 1$ with sigmoid activation; maximum sensor correction $0.5$ \\

Dropout
& Projection dropout $0.2$; regressor dropout $0.3$
& Regressor dropout $0.3$
& Visual projection dropout $0.2$; Transformer/regressor dropout $0.3$; sensor-fusion dropout $0.25$ \\
\bottomrule
\end{tabular}
\end{table}

\begin{table}[H]
\centering
\small
\caption{Training hyperparameters used across the temporal models.}
\label{tab:training-hyperparameters}
\begin{tabular}{ll}
\toprule
Hyperparameter & Value \\
\midrule
Sequence length & $10$ seconds \\
Optimizer & Adam \\
Loss & Weighted MSE \\
Attention-bin weighting factor & $\alpha = 0.5$ \\
Learning rate, visual GRU & $10^{-4}$ \\
Learning rate, sensor GRU & $5 \times 10^{-4}$ \\
Learning rate, fusion and fairness models & $10^{-4}$ \\
Weight decay & $10^{-4}$ \\
Batch size, visual GRU & $32$ \\
Batch size, sensor GRU & $128$ \\
Batch size, fusion and fairness models & $32$ \\
Early stopping patience & $7$ epochs \\
Learning-rate scheduler & ReduceLROnPlateau \\
Scheduler factor & $0.3$ \\
Scheduler patience & $3$ epochs \\
Gradient clipping & Maximum norm $1.0$ \\
Training seeds & $10$ independent seeds \\
\bottomrule
\end{tabular}
\end{table}
\FloatBarrier

The 10 random seeds used for all baseline and fairness-aware experiments were
\(42, 100, 2000, 2025, 2026, 2027, 2048, 4096, 7000, 8192\).

\FloatBarrier
\clearpage
\end{document}